\documentclass[10pt]{article}
\usepackage[tbtags]{amsmath}
\usepackage[linktocpage]{hyperref}
\hypersetup{colorlinks, linkcolor={red!60!black}, citecolor={blue!60!black}, urlcolor={blue!80!black}}
\usepackage{amssymb,amsthm,mathrsfs}
\usepackage{amsfonts}
\usepackage{graphicx}
\usepackage{stackrel}
\usepackage{subcaption}
\graphicspath{{figs/}}
\usepackage[left=2.5cm,right=2.2cm,top=0.5cm,bottom=1.3cm,includeheadfoot]{geometry}
\usepackage{indentfirst}
\usepackage{color}
\usepackage{cite}
\usepackage{mathtools}
\usepackage{tikz-cd}
\usepackage{enumitem}
\usepackage{multicol}
\usepackage{float}

\usepackage[normalem]{ulem}

\newcommand{\be}{\begin{equation}}
\newcommand{\ee}{\end{equation}}
\newcommand{\de}{\mbox{d}}

\newcommand{\coupling}{\kappa}

\numberwithin{equation}{section}

\renewcommand*{\thefootnote}{\fnsymbol{footnote}}
\begin{document}
	\begin{center}
		{\Large\bf
		Dipolar perturbations of nonbidiagonal black holes in bigravity
		}
		\vskip 5mm
		{\large
			David Brizuela$^{1,}$\footnote{e-mail address: {\tt david.brizuela@ehu.eus}},
			Marco de Cesare$^{2,3,}$\footnote{e-mail address: {\tt marco.decesare@na.infn.it}}, and Araceli Soler Oficial$^{1,}$\footnote{e-mail address: {\tt araceli.soler@ehu.eus}}}
		\vskip 3mm
		{\sl $^{1}$Department of Physics and EHU Quantum Center, University of the Basque Country UPV/EHU,\\
			Barrio Sarriena s/n, 48940 Leioa, Spain}\\\vskip 1mm
		{\sl $^2$ Scuola Superiore Meridionale, Largo San Marcellino 10, 80138 Napoli, Italy}\\\vskip 1mm
        {$^3$ INFN, Sezione di Napoli, Italy}
	\end{center}
	
\setcounter{footnote}{0}
\renewcommand*{\thefootnote}{\arabic{footnote}}
 
	\begin{abstract} 
In bimetric gravity, nonbidiagonal solutions describing a static, spherically symmetric, and asymptotically flat black hole are given by a pair of Schwarzschild geometries, one in each metric sector. The two geometries are linked by a nontrivial diffeomorphism, which can be fully determined analytically if the two geometries possess the same isometries. This exact solution depends on four free parameters: the mass parameters of the two black holes, the ratio between the areal radii of the two metrics, and the proportionality constant between their (appropriately normalized) time-translation invariance Killing vector fields. We study the dynamics of axial dipolar perturbations on such a background and obtain general analytical solutions for their evolution. We show that, in general, the characteristic curves followed by dipolar gravitational waves are spacelike with respect to both metrics, and thus the propagation is superluminal. In fact, the velocity of a pulse, as measured by a static observer, turns out to increase with the distance to the black hole. The only exception to this general behavior corresponds to the special case where the two proportionality constants linking the areal radii and the Killing vectors coincide, for which waves travel at the speed of light. Therefore, we conclude that this is the only physically reasonable background, and thus our results restrict the class of viable black-hole solutions in bimetric gravity.
      \end{abstract}

\section{Introduction}

Bimetric gravity \cite{Hassan:2011vm, Hassan:2011tf} is a consistent nonlinear theory of gravity that extends the framework of general relativity (GR) by introducing two interacting dynamical metrics. The interactions between the two metrics are governed by a specifically constructed potential designed to eliminate the Boulware-Deser ghost, necessary to ensure the consistency of the theory. The bimetric framework provides a richer dynamical structure
than GR, enabling the exploration of new solutions to fundamental questions in cosmology. For example, within this theory, it is possible to address the accelerated expansion of the Universe without invoking a cosmological constant or exotic dark-energy components \cite{Volkov:2011an, Volkov:2012zb, DeFelice:2014nja, Akrami:2015qga}. Furthermore, the theory features an extra massive spin-2 mode, in addition to the standard massless spin-2 mode of GR, which arises due to the presence of the second metric, and may represent a natural candidate for dark matter \cite{Aoki:2014cla, Bernard:2014psa, Blanchet:2015sra, Blanchet:2015bia}. However, in the regime where bimetric gravity is able to produce a viable dark-matter particle, the theory is expected to be indistinguishable from GR \cite{Babichev:2016hir}. Moreover, Ref.~\cite{Dwivedi:2024okk} has shown that the theory may potentially alleviate the Hubble tension. Observational analyses \cite{vonStrauss:2011mq, Caravano:2021aum, Hogas:2021lns, Hogas:2021saw} and theoretical studies \cite{Hogas:2021fmr} have placed constraints on the parameter space of the theory, reinforcing its viability. An exhaustive analysis of constraints from local observations, cosmological data, gravitational wave studies, and theoretical stability requirements can be found in Ref.~\cite{Hogas:2022owf}. 

Moving beyond cosmological spacetimes, to assess the viability of bimetric gravity as an alternative to general relativity, the theory should also withstand phenomenological tests in the context of black-hole spacetimes. For this reason, the stability and viability of black-hole solutions in bimetric gravity have been widely studied in the literature \cite{Babichev:2015xha, Torsello:2017cmz}. The simplest solutions one can consider are static and spherically symmetric, which can be divided into two different branches \cite{Comelli:2011wq}. The first branch corresponds to the so-called \textit{bidiagonal solutions}, where it is possible to find a coordinate system in which the two metrics can be simultaneously diagonalized. The second branch corresponds to \textit{nonbidiagonal solutions}, where such a coordinate system does not exist.

Regarding bidiagonal solutions, all exact static black-hole solutions with no charge that have been found in spherical symmetry correspond to the standard GR solutions (i.e., Schwarzschild, Schwarzschild-de Sitter, and Schwarzschild-anti-de Sitter) \cite{Babichev:2014tfa, Babichev:2014fka}.  However, dynamical stability analyses of linear perturbations around such bidiagonal bi-Schwarzschild black holes have shown that such solutions are unstable \cite{Babichev:2013una, Brito:2013wya}. This instability was later confirmed in Ref.~\cite{Babichev:2014oua}, and Ref.~\cite{Torsello:2017cmz} later showed the Lyapunov instability of these solutions. These results do not apply to numerical solutions describing hairy bidiagonal black holes \cite{Brito:2013xaa}, which have been found to be stable within certain parameter regions \cite{Gervalle:2020mfr}. Interestingly, static and spherically symmetric nonbidiagonal background solutions do not exhibit unstable modes under radial perturbations \cite{Babichev:2014oua}. Studies of quasinormal modes have further reinforced these results, highlighting the differences in the stability properties of exact bidiagonal and nonbidiagonal cases \cite{Babichev:2015zub}. Analogous results have been obtained in the context of ghost-free multimetric gravity  \cite{Wood:2024acv, Wood:2024eol}.

In fact, nonbidiagonal solutions have some interesting physical properties: for vacuum, static, spherically symmetric configurations, both metric sectors satisfy the vacuum Einstein equations with effective cosmological constants \cite{Comelli:2011wq}, and have collinear time-translation invariance Killing vector fields. As a result, they correspond to Schwarzschild-(anti)de Sitter geometries, and thus, deviations from GR emerge only at the perturbative level. In these cases, the analytical form of the background solution can be determined up to a function that satisfies a nonlinear partial differential equation \cite{Volkov:2012wp, Volkov2015} and, in general, even if the isometry groups of both metrics are isomorphic, their isometries may differ \cite{Torsello:2017ouh}. However, if, additionally, one requires both metrics to be asymptotically flat and to have the same isometries
(so that every isometry of one metric is also an isometry of the other), the most general nonbidiagonal solution turns out to be unique up to four constants: the two mass parameters (one for each metric), the ratio between the areal radii of the two metrics, and the proportionality constant between their corresponding (appropriately normalized) time-translation invariance Killing vector fields \cite{Brizuela:2024smr}. All previously mentioned works on nonbidiagonal solutions assume a particular value for this last constant, and our purpose is to study the physical implications in the general case. 

The main goal of this paper is to study the explicit evolution of the perturbations on general spherically symmetric and asymptotically flat nonbidiagonal black-hole backgrounds. We will restrict ourselves to the simplest possible case: dipolar waves with axial (sometimes also known as odd) polarity. However, this will be enough to heavily restrict the physically reasonable values of the free parameters. In general, we will find that the characteristic curves followed by  dipolar gravitational waves are spacelike, which implies a superluminal propagation. Only when the two commented background parameters (the ratio between the areal radii of the two metrics and the proportionality constants linking the Killing vector fields) coincide, the waves propagate at the speed of light and the solution is compatible with known physics.

The remainder of the paper is organized as follows. In Sec.~\ref{Sec: background} we introduce the nonbidiagonal background geometry we are going to work with, ensuring that it is smooth and well-defined. Once we have the background, in Sec.~\ref{Sec: axial-perturbations}, we study the propagation of axial dipolar gravitational waves, analyzing their physical properties for the different values of the parameters of the theory. In Sec.~\ref{Sec: conclusions}, we summarize the main results of our study and discuss their implications. Finally, in the \hyperlink{appendix}{Appendix} we present the configurations of the light cones for the background under consideration.

\section{Background geometry}\label{Sec: background}

The bimetric gravity theory developed by Hassan and Rosen \cite{Hassan:2011zd} considers two interacting dynamical metrics, $g$ and $f$, on a four-dimensional spacetime manifold. The two metrics are nonlinearly coupled, and the action of the theory is
\be\label{Eq:BimetricAction}
S_{\rm\scriptscriptstyle HR}=\frac{M_g^2}{2} \int \de^4 x \sqrt{- \det{g}}\, {\cal R}^{(g)}
+\frac{M_f^2}{2} \int \de^4 x \sqrt{-\det{f}}\, {\cal R}^{(f)}-\bar{m}^2 M_g^2 \int \de^4x \sqrt{-\det{g}}\,\sum_{n=0}^4 \beta_n e_n(\mathbb{S})~,
\ee
where ${\cal R}^{(g)}$ and ${\cal R}^{(f)}$ are the Ricci scalars of the metrics $g$ and $f$, respectively. The coupling constants $M_g$, $M_f$, and $\bar{m}$ have dimensions of mass, while the $\beta_n$ are dimensionless. The terms $e_n$ are the elementary symmetric polynomials constructed from scalar combinations of the matrix $\mathbb{S}\coloneqq \sqrt{g^{-1}f}$, as defined in Refs.~\cite{Hassan:2011hr,Bernard:2015mkk}. Matter fields are typically assumed to couple only to one of the metrics \cite{deRham:2014naa, Yamashita:2014fga, Schmidt-May:2015vnx}, which in our case is assumed to be $g$ and thus we will refer to it as the physical metric. The background equations in spherical symmetry can be found in Ref.~\cite{Brizuela:2024smr}, whose notation and conventions we follow.

As shown in Refs.~\cite{Comelli:2011wq, Volkov2015}, if one assumes, on the one hand, a spherically symmetric and static metric $g$, and, on the other hand, that there is no common chart where both $f$ and $g$ can be simultaneously diagonalized, the equations of motions for the bimetric theory reduce to two uncoupled copies of the vacuum Einstein equations, one for each metric sector. Therefore, the general solution for both $f$ and $g$ metrics is the Schwarzschild-(anti)de Sitter geometry, parametrized by their corresponding mass parameters, $\mu_f$ and $\mu_g$, and effective cosmological constants, $\Lambda_f$ and $\Lambda_g$ determined by the parameters of the theory. For the subsequent analysis, we will also assume asymptotic flatness for both geometries,\footnote{Since the two metrics are defined on the same manifold, topological consistency demands that they have the same asymptotic structure  \cite{Torsello:2017ouh}.} and therefore impose the conditions $\Lambda_g=\Lambda_f=0$, and that both mass parameters are positive $\mu_f>0$ and $\mu_g>0$.

In general, we will work in the Schwarzschild coordinates $(t,r)$ of the metric $g$, which are valid either outside $(r\in(2\mu_g,\infty)\,)$ or inside $(r\in(0,2\mu_g)\,)$ the horizon of $g$. Similarly, the geometry of $f$ also has a horizon, located at $r=2\mu_f$.\footnote{Note that the definition of the integration constant $\mu_f$ used in this work differs from that in Ref.~\cite{Brizuela:2024smr}.} The presence of the horizons splits the spacetime into three different domains, described by a corresponding range of $r$. More precisely, there is a region $A$ interior to both horizons with $0<r<{\rm min}\{2\mu_f,2\mu_g\}$, an intermediate region $B$ between horizons with ${\rm min}\{2\mu_f,2\mu_g\}<r<{\rm max}\{2\mu_f,2\mu_g\}$, and an exterior region $C$ for $r>{\rm max}\{2\mu_f,2\mu_g\}$. In this way, in each of these regions, the line elements read,
\begin{subequations}\label{Eq:background-line-elements}
\begin{align}
   \text{d}s_g^2 &=-\Sigma_g\text{d}t^2+\Sigma_g^{-1}\text{d}r^2+r^2\left(\text{d}\theta^2+\sin^2{\theta}\text{d}\phi^2\right)~,\label{Eq:metric-g}\\
    \text{d}s_f^2&=-\Sigma_f\dot{T_I}^2\text{d}t^2-2\Sigma_f\dot{T_I}T_I^\prime\text{d}t\text{d}r+\left(\omega^2\Sigma_f^{-1}-\Sigma_fT_I^\prime{}^2\right)\text{d}r^2+\omega^2r^2\left(\text{d}\theta^2+\sin^2{\theta}\text{d}\phi^2\right)~,\label{Eq:metric-f}
    \end{align}
\end{subequations}
where we have defined $\Sigma_g=1-2\mu_g/r$ and $\Sigma_f=1-2\mu_f/r$, while the subscript $I=A,\,B,\,C$ denotes the region. The function $T_I=T_I(t,r)$ is implicitly defined by the differential equation
\begin{equation}\label{eq:Tequation}
T^{\prime\,2}_I=\left(\frac{1}{\Sigma_g}-\frac{1}{\Sigma_f} \right)\left(\frac{\Dot{T_I}^2}{\Sigma_g}-\frac{\omega^2}{\Sigma_f}\right),
\end{equation}
where a dot and a prime denote derivatives with respect to $t$ and $r$, respectively, and $\omega$ is a positive constant that measures the ratio between the areal-radius functions of each metric. In fact, this equation provides the coordinate transformation from $(t,r)$ to the Schwarzschild coordinates $(T_I,\omega r)$ of the metric $f$. In particular, we note that the assumption of nonbidiagonal solutions implies that neither $\dot{T}_I$ nor $T_I'$ can vanish identically.

Therefore, while $\partial_{t}$ is the Killing vector field of the metric $g$ that encodes time-translation invariance outside its horizon and homogeneity inside its horizon, $\partial_{T_I}=\frac{1}{\dot{T}_I}\partial_t$ corresponds to such Killing field for the metric $f$. Note that, although the vector $\partial_{T_I}$ defines the same integral curves as $\partial_t$, generically $\partial_{T_I}$ is not a Killing field of the metric $g$---unless $\dot{T}_I$ is a constant. That is, both these Killing vectors generate the same isometry, and thus both metrics have the same full set of isometries (so that every isometry of one metric is also an isometry of the other), if and only if $\dot{T}_I$ is a constant. In theories with only one metric, such as GR, the Killing vector fields are defined up to an overall constant, which is not physically meaningful. In particular, the above Killing fields have been normalized so as to have unit norm at spatial infinity as computed with their corresponding metric, i.e., asymptotically  $g(\partial_t,\partial_t)\to -1$ and  $f(\partial_{T_I},\partial_{T_I})\to -1$. However, in a theory like the present one, with two metrics, even if it is a constant, the relative norm between the Killing fields, as measured by any of the metrics, that is, $g(\partial_t,\partial_t)/g(\partial_{T_I},\partial_{T_I})=f(\partial_t,\partial_t)/f(\partial_{T_I},\partial_{T_I})=\dot{T}_I^2$ does have physical consequences, as will be explicitly shown below.

In the context of the above general solution, we will henceforth assume that both metrics have the same isometries. As explained above, this implies that $\dot{T_I}\coloneqq c$, with $c$ an arbitrary constant. Under this assumption, it is easy to check that, in a given region $I$, Eq.~\eqref{eq:Tequation} admits the exact solution
\begin{equation}\label{Eq:solTquadrature}
 T_I(t,r) =c\, t+\sigma_{\scriptscriptstyle I}\int \frac{dr}{|\Sigma_g\Sigma_f|}\sqrt{\frac{2}{r}(\mu_g-\mu_f)W}~,
\end{equation}
where, for convenience, we have defined
\begin{equation}\label{Eq:DefineW}
 W:=c^2-\omega^2-\frac{2}{r}(c^2\mu_f-\omega^2\mu_g)~,
 \end{equation}
and $\sigma_{\scriptscriptstyle I}=\pm 1$ is an arbitrary sign factor. We note that this factor $\sigma_I$ is independent of the sign of $c$, and that, at this stage, it can be chosen independently in each of the above-defined regions $A$, $B$, and $C$.

Next, to ensure the existence of a real solution for $T_I$ for all the regions, we must require that the square-root argument in the integrand in \eqref{Eq:solTquadrature} is positive for all $r>0$. We note also that we must exclude the case with an identically vanishing square root in \eqref{Eq:solTquadrature}, as it would imply $T_I'=0$ identically and thus a bidiagonal solution. In particular, this indicates that the two black-hole horizons cannot coincide, that is $\mu_g\neq\mu_f$.

On the one hand, in the special case $|c|=\omega$, the square-root argument is a perfect square, and thus the existence of the nonbidiagonal solution is guaranteed for any values of the masses, as long as $\mu_g\neq\mu_f$. On the other hand, for $c^2\mu_f=\omega^2\mu_g$, the function $W$ is a constant and the sign of the square-root argument reduces to the sign of $(\mu_g-\mu_f)(c^2-\omega^2)$. However, if $|c|\neq\omega$ and $c^2\mu_f\neq\omega^2\mu_g$, the square-root argument does not have a definite sign for generic values of the parameters. Specifically, the function $W$ has a simple pole at the origin $r=0$, where it changes sign, and, apart from that, it has one, and only one, real root at $r=2(c^2\mu_f-\omega^2\mu_g)/(c^2-\omega^2)\neq 0$. Therefore, demanding that this root be negative, implies that $W$ has a definite sign for all $r>0$. It turns out that this requirement alone, without any further restrictions on the parameters, automatically implies that ${\rm sgn}(W)={\rm sgn}(\mu_g-\mu_f)$, which provides a positive sign for the square-root argument. This also excludes bidiagonality at isolated points.

In this way, one can identify three different cases that provide a real solution for $T_I$ \eqref{Eq:solTquadrature}, such that $T_I'\neq0$ for all $r>0$:
\begin{itemize}
\item $\omega<|c|$ and $c^2\mu_f\leq \omega^2\mu_g$ $\Longrightarrow{\rm sgn}(W)={\rm sgn}(\mu_g-\mu_f)=1$~,
\item $|c|<\omega$ and $\omega^2\mu_g \leq c^2\mu_f$ $\Longrightarrow{\rm sgn}(W)={\rm sgn}(\mu_g-\mu_f)=-1$~,
\item $|c|=\omega$ and $\mu_g\neq\mu_f$\qquad\hspace{0.01cm}
$\Longrightarrow$ ${\rm sgn}(W)={\rm sgn}(\mu_g-\mu_f)$~,
\end{itemize}
with $\mu_g$, $\mu_f$, and $\omega$ positive. Therefore, in the first case, the function $W$ is positive definite for all $r>0$ and one has $\mu_f<\mu_g$, which implies that the horizon of the metric $g$ lies outside that of the metric $f$, while in the second case, $W$ is negative definite, $\mu_g<\mu_f$, and the configuration is reversed with the horizon of the metric $g$ lying inside that of the metric $f$. In both cases, asymptotically one has $W\to (c^2-\omega^2)$ as $r\to+\infty$. Finally, as in the previous two cases, in the third one the sign of $W$ is fixed by the difference between the mass parameters, ${\rm sgn}(W)={\rm sgn}(\mu_g-\mu_f)$; however, unlike the previous cases, both configurations, either $\mu_g<\mu_f$ or $\mu_g>\mu_f$, are allowed. We note that, in any of the cases listed above, one has $\mu_f\neq\mu_g$. From now on, we will assume that the background parameters fall within one of the above-listed cases.

Once the existence of the solution is guaranteed, we need to ensure its smoothness. In particular, it can be readily realized from \eqref{Eq:solTquadrature} that both $T_I$ and $T'_I$ diverge as $r$ approaches either black-hole horizon $\Sigma_g=0$ or $\Sigma_f=0$. Since neither $T_I$ nor its derivatives represent physical fields, their divergence per se does not imply an ill-posedness of the theory. However, we need to require that physical fields are smooth. Thus, we will impose continuity of the components of the metric $f$ in the domain of the chart $(t,r)$ (as specified, in turn, by the components of the metric $g$), and similarly for the components of the metric $g$ in the domain of the charts $(T_{I},r)$. This condition will lead to an appropriate choice of the sign factors $\sigma_{I}$. More precisely, on the one hand, when the metric $f$ is expressed in the $(t,r)$ chart, as in Eq.~\eqref{Eq:metric-f}, its components read
\begin{subequations}
\begin{align}
 f_{tt}&=-\Sigma_f c^2~,\\
 f_{tr}&=-\sigma_{\scriptscriptstyle\rm I}\, c \frac{{\rm sgn}(\Sigma_f) }{|\Sigma_g|}\sqrt{\frac{2}{r}(\mu_g-\mu_f)W}~,\\
 f_{rr}&=\frac{c^2+\omega^2}{\Sigma_g}-\frac{c^2\Sigma_f}{\Sigma_g^2}~.
 \end{align}
\end{subequations}
We note that $f_{tr}$ contains a potential discontinuity at $\Sigma_f=0$, while the remaining metric components are smooth there. In order to avoid such a discontinuity, it is sufficient to ensure that $f_{tr}$ keeps the same sign on either side of the $\Sigma_f=0$ surface. On the other hand, concerning the metric $g$, when expressed in the $(T_{I},\,r)$ chart, its components read
\begin{subequations}
\begin{align}
 g_{T_{\scriptscriptstyle\rm I}T_{\scriptscriptstyle\rm I}}&=-\frac{\Sigma_g}{ c^2}~,\\
 g_{T_{\scriptscriptstyle\rm I}r}&=\sigma_{\scriptscriptstyle\rm I} \frac{{\rm sgn}(\Sigma_g) }{c^2|\Sigma_f|}\sqrt{\frac{2}{r}(\mu_g-\mu_f)W}~,\\
 g_{rr}&=\frac{c^2+\omega^2}{c^2\Sigma_f}-\frac{\omega^2\Sigma_g}{c^2\Sigma_f^2}~.
 \end{align}
\end{subequations}
While $ g_{T_{\scriptscriptstyle\rm I}T_{\scriptscriptstyle\rm I}}$ and $g_{rr}$ are smooth everywhere outside the horizon $\Sigma_f=0$, where the chart $(T_I,r)$ breaks down, depending on $\sigma_I$, the component $g_{T_{\scriptscriptstyle\rm I}r}$ may be discontinuous at $\Sigma_g=0$. Therefore, in order to ensure the continuity of $f_{tr}$ at $r=2\mu_f$ and of $g_{T_{\scriptscriptstyle\rm I}r}$ at $r=2\mu_g$, we will require $\sigma_I={\rm sgn}(\Sigma_f\Sigma_g)\big\lvert_I$~, where the sign function on the right-hand side is evaluated in the $I$th region.

To summarize, fixing the sign factors as shown above so that physical fields are continuous in the domain of the corresponding chart, the solution for $T_I$ reads 
\begin{equation}\label{Eq:T_finalform}
    T_I(t,r) =c\, t+\int dr\, \frac{ 1}{\Sigma_f \Sigma_g}\sqrt{\frac{2}{r}(\mu_g-\mu_f)W}~,
\end{equation}
which implies that ${\rm sgn}(T')={\rm sgn}(\Sigma_f \Sigma_g)$, while ${\rm sgn}(\dot{T})={\rm sgn}(c)$. Therefore, the background solution \eqref{Eq:background-line-elements}, with the function $T_I$ given by \eqref{Eq:T_finalform}, is completely characterized by four parameters (the two mass parameters $\mu_g$ and $\mu_f$ that define their corresponding horizons, the ratio between the areal-radius functions $\omega$, and the proportionality constant $c$ between the Killing vectors $\partial_t$ and $\partial_{T_I}$), which must obey the set of relations presented above. While $\mu_g$, $\mu_f$, and $\omega$ are positive by definition, the sign of $c$ determines whether $\partial_t$ and $\partial_{T_I}$ are parallel or antiparallel. From now on, we will use the chart $(t,r)$, which is well-defined both outside and inside the horizon of the metric $g$. In particular, we note that, at the horizon of the metric $f$, where $\Sigma_f=0$, even if $f_{tt}$ vanishes, $f_{tr}$ and $f_{rr}$ are regular and nonvanishing, ${\rm det}(f)\neq 0$, and thus in this chart the metric $f$ is nondegenerate at its horizon. The above conditions also ensure that the matrix $\mathbb{S}$ is well-behaved, cf.~Ref.~\cite{Brizuela:2024smr}. For completeness, and to conclude the analysis of the background, in the \hyperlink{appendix}{Appendix} we examine the possible configurations of the light cones of the two metric sectors.

For later purposes, it is convenient to define the auxiliary functions,
\begin{subequations}
\begin{align}
     &P(r)\coloneqq\frac{c^2(\Sigma_f-\Sigma_g)}{\Sigma_g^2}=\frac{2 c^2 (\mu_g -\mu_f)r}{  (r-2\mu_g)^2}~,\label{Eq:DefineU}\\
      &Q(r)\coloneqq\frac{c}{\Sigma_g}\,\sqrt{{(\Sigma_f-\Sigma_g)(c^2\Sigma_f-\omega^2\Sigma_g)}}=   \frac{c\sqrt{ 2(\mu_g  -\mu_f)r W(r)}}{(r-2 \mu_g)}~,\label{Eq:DefineV}
   \end{align}
\end{subequations}
the combinations of the coupling constants that appear in the action \eqref{Eq:BimetricAction},
\begin{align}
\alpha &:=\frac{M_f}{M_g},\\
\coupling &:=\frac{(\beta_1+\omega\beta_2)\bar{m}^2}{\omega^2(|c|+\omega)},
\end{align}
as well as the sign $s:={\rm sgn}(Q/W)$, which, for the configuration of parameters detailed above, is given in terms of $c$ and the difference between the horizon radii, that is, $s=\, {\rm sgn}(c(\mu_g-\mu_f))$. This sign will turn out to be of key relevance to determine the properties of the propagation of the waves on this background.

\section{Axial dipolar perturbations}\label{Sec: axial-perturbations}

In this section we consider axial dipolar perturbations propagating on the background described above. This section is divided into two subsections. In Sec.~\ref{Subsec: solution} the linearized equations of motion are presented and the corresponding general solution is derived. Then, in Sec.~\ref{Subsec: properties} the physical properties of the solution are analyzed, both in the dynamical and static cases.

\subsection{Equations of motion and general solution}\label{Subsec: solution}

The equations for linear perturbations around the most general spherically symmetric background within bimetric gravity have been derived in Ref.~\cite{Brizuela:2024smr}. The perturbations of the background metrics $g$ and $f$, are encoded in two symmetric rank-two tensor fields $h_{(g)}$ and $h_{(f)}$, respectively. Since the background is assumed to be spherically symmetric, it is convenient to decompose such perturbations in tensor spherical harmonics. Here, we will only consider dipolar (i.e., $l=1$) perturbations with axial (sometimes also called odd) polarity. In this way, the components of $h_{(g)}$ and $h_{(f)}$ in the $(t,r)$ chart read
\begin{equation}
   h_{(i)}= \begin{pmatrix}
        0 & 0 & h_{(i)0}^m\csc{\theta}\partial_\phi Y^m_1 & -h_{(i)0}^m\sin{\theta}\partial_\theta Y^m_1 \\
        \ast & 0 & h_{(i)1}^m\csc{\theta}\partial_\phi Y^m_1 & -h_{(i)1}^m\sin{\theta}\partial_\theta Y^m_1 \\
        \ast & \ast & 0 & 0 \\
        \ast & \ast & \ast & 0
    \end{pmatrix},
\end{equation}
where $Y^m_l(\theta,\phi)$ are the usual spherical harmonics, $m=-1,0,1$, and the index $i=g,f$ denotes the metric sector. In the axial sector with $l=1$, there is one gauge degree of freedom and the four components $\{h_{(g)0}^{m}(t,r),\,h_{(g)1}^{m}(t,r),\,h_{(f)0}^{m}(t,r),\,h_{(f)1}^{m}(t,r)\}$  describe one dynamical propagating degree of freedom (see Ref.~\cite{Brizuela:2024smr}). This is quite different from GR, where the axial dipolar mode is nonpropagating. Taking into account that the equations of motion are independent of $m$ and, in order to  make the notation lighter, in the following we remove the harmonic label $m$ from the different variables. Similarly, the functional dependence on $(t,r)$ is implied.

Assuming the nonbidiagonal background defined by \eqref{Eq:metric-g}, \eqref{Eq:metric-f}, and \eqref{Eq:T_finalform}, the linearized equations for the axial sector with $l=1$ read \cite{Brizuela:2024smr}
\begin{subequations}
   \begin{align}
h^{\prime\prime}_{(g)0}-\frac{2}{r}\dot{h}_{(g)1}-\dot{h}_{(g)1}^\prime-\frac{2}{r^2}h_{(g)0}+\coupling\left((h_{(f)0}-\omega^2h_{(g)0})P-(h_{(f)1}-\omega^2h_{(g)1})Q\right)=0~, \label{Eq:g-primeprime} \\
\Ddot{h}_{(g)1}-\dot{h}^{\prime}_{(g)0}+\frac{2}{r}\dot{h}_{(g)0}-\coupling\left((h_{(f)0}-\omega^2h_{(g)0})Q-(h_{(f)1}-\omega^2h_{(g)1})W\right)=0~,\label{Eq:g-dotdot} \\
h^{\prime\prime}_{(f)0}-\frac{2}{r}\dot{h}_{(f)1}-\dot{h}^\prime_{(f)1}-\frac{2}{r^2} h_{(f)0}-\frac{|c|\,\coupling}{\alpha^2\omega}\left((h_{(f)0}-\omega^2h_{(g)0})P-(h_{(f)1}-\omega^2h_{(g)1})Q\right)=0~,\label{Eq:f-primeprime} \\
\Ddot{h}_{(f)1}-\dot{h}^{\prime}_{(f)0}+\frac{2}{r}\dot{h}_{(f)0} + \frac{|c|\,\coupling}{\alpha^2\omega}\left((h_{(f)0}-\omega^2h_{(g)0})Q-(h_{(f)1}-\omega^2h_{(g)1})W\right)=0~,\label{Eq:f-dotdot}
\end{align}
\end{subequations}
with $W$, $P$, and $Q$ defined in Eqs.~\eqref{Eq:DefineW}, \eqref{Eq:DefineU}, and \eqref{Eq:DefineV}, respectively.

These equations are not all independent, due to the existence of the Bianchi constraints\footnote{Following the notation in Ref.~\cite{Brizuela:2024smr}, the Bianchi constraints read $2 t^{(i)}{}^Av^{(i)}_A+\overset{(i)}{\nabla}{}_At^{(i)}{}^A=0,$ for $i=g,\,f$.}. For this reason, let us introduce the gauge-invariant variables $ \Pi_{(i)}\coloneqq 2h_{(i)0}-rh_{(i)0}^{\prime}+r\dot{h}_{(i)1}$, for $i=g,\,f$, along with $h_{(-)0}\coloneqq h_{(f)0}-\omega^2h_{(g)0}$ and  $h_{(-)1}\coloneqq h_{(f)1}-\omega^2h_{(g)1}$. Up to a rescaling, these definitions parallel those given in Ref.~\cite{Gerlach:1979rw} in the case of GR for general  $l$-modes, and have also been employed in bigravity in Ref.~\cite{Babichev:2015zub}. Note that the four variables $(\Pi_{(g)},\Pi_{(f)},h_{(-)0},h_{(-)1})$ are not all independent, since they are subject to the following identity:
\begin{equation}\label{Eq:Pi_diff_identity}
    \Pi_{(f)}-\omega^2\Pi_{(g)}=2h_{(-)0}-rh_{(-)0}^{\prime}+r\dot{h}_{(-)1}\,,
\end{equation}
which follows directly from the definitions above. In particular, from these new variables, one can recover three out of four components of the metric perturbations, which makes the gauge degree of freedom explicit. For instance, one can write,
\begin{subequations}\label{Eq:solution-metric-perturbations}
\begin{align}
h_{(g)0}&=r^2 C(t)+r^2\int\left(\frac{\dot{h}_{(g)1}}{r^2}-\frac{\Pi_{(g)}}{r^3}\right)\de r~,\label{Eq:solutionh0g}\\
h_{(f)0}&=h_{(-)0}+\omega^2h_{(g)0}~,\label{Eq:solutionh0f}\\
h_{(f)1}&=h_{(-)1}+\omega^2h_{(g)1}~,
\end{align}
\end{subequations}
 where $h_{(g)1}$ is chosen to parametrize the gauge freedom, and $C(t)$ is a free integration function that must be fixed consistently with boundary conditions.

Let us now solve the system of equations \eqref{Eq:g-primeprime}-\eqref{Eq:f-dotdot} for the new four variables, which reads
\begin{subequations}
    \begin{align}
    \frac{\Pi_{(g)}^\prime}{r}+\frac{\Pi_{(g)}}{r^2}-\coupling\left(P h_{(-)0}-Q h_{(-)1}\right)=0~,\\
    \frac{\dot{\Pi}_{(g)}}{r}-\coupling\left(Q h_{(-)0}-W h_{(-)1}\right)=0~,\label{Eq:Pigdot}\\
    \frac{\Pi_{(f)}^\prime}{r}+\frac{\Pi_{(f)}}{r^2}+\frac{|c|\,\coupling}{\alpha^2\omega}\left(P h_{(-)0}-Q h_{(-)1}\right)=0~,\\
    \frac{\dot{\Pi}_{(f)}}{r}+\frac{|c|\,\coupling}{\alpha^2\omega}\left(Q h_{(-)0}-W h_{(-)1}\right)=0~.
    \end{align}
\end{subequations}
From here, we observe that upon defining the auxiliary quantity $\Pi_{(+)}\coloneqq\Pi_{(f)}+\frac{|c|}{\alpha^2\omega}\Pi_{(g)} $, the system can be recast in the simpler form
\begin{subequations}
\begin{align}
    &\Pi_{(+)}^\prime+\frac{\Pi_{(+)}}{r}=0~,\label{Eq:Pi+prime}\\
    &\dot{\Pi}_{(+)}=0~,\label{Eq:Pi+dot}\\
    &\Pi_{(g)}^\prime+\frac{\Pi_{(g)}}{r}-\frac{Q}{W}\dot{\Pi}_{(g)}=0~,\label{Eq:Pigprime}\\
    &h_{(-)1}=\frac{Q}{W}h_{(-)0}-\frac{1}{\coupling}\frac{\dot{\Pi}_{(g)}}{r\,W}~.\label{Eq:h1minus}
    \end{align}
\end{subequations}
Moreover, it follows from the identity \eqref{Eq:Pi_diff_identity} and \eqref{Eq:Pigdot} that
\begin{equation}
    \Pi_{(f)}-\omega^2\Pi_{(g)}=2 h_{(-)0}-r h_{(-)0}^{\prime}+r\frac{Q}{W}\dot{h}_{(-)0}-\frac{1}{\coupling}\frac{\Ddot{\Pi}_{(g)}}{W}~.\label{Eq:h0minus}
\end{equation}
We note that Eq.~\eqref{Eq:Pigprime} for $\Pi_{(g)}$, as well as the Eqs.~\eqref{Eq:Pi+prime} and \eqref{Eq:Pi+dot} for $\Pi_{(+)}$, are decoupled from the rest of the variables. On the one hand, Eqs.~\eqref{Eq:Pi+prime}--\eqref{Eq:Pi+dot} have the same structure as the constraint equations for a dipolar perturbation in vacuum general relativity, and can be easily solved,
\begin{equation}
 \Pi_{(+)}=\frac{K}{r}~,
\end{equation}
with a constant $K$. On the other hand, \eqref{Eq:Pigprime} describes a propagating perturbation and its solution reads
\begin{equation}
 \Pi_{(g)}(t,r)=\frac{F_1(t+R(r))}{r}~,
\end{equation}
where $F_1(t+R(r))$ is a free function and we have defined,
\begin{equation}\label{Eq:integral-I}
    R(r):= \int{\frac{Q(r)}{W(r)}\text{d}r}=
    \left\{\begin{matrix}
    s\,\left(r+2\mu_g\ln{|\frac{r}{2\mu_g}-1|}\right)~,\quad\text{for}\quad  |c|=\omega~,\\[.5cm]
\frac{2 c\sqrt{2(\mu_g-\mu_f)r\,W(r)}}{c^2-\omega^2}+2\mu_g\,\ln{\left|\frac{\sqrt{r\,(\mu_g-\mu_f)W(r)}-\sqrt{2} c(\mu_g-\mu_f)}{\sqrt{r\,(\mu_g-\mu_f)W(r)}+\sqrt{2} c(\mu_g-\mu_f)}\right|}~,\quad\text{for}\quad  |c|\neq\omega~,
\end{matrix}
\right.
\end{equation}
with the sign $s={\rm sgn}(c(\mu_g-\mu_f))$. In particular, note that, as commented above, for the configuration of the parameters under consideration, ${\rm sgn}(W(r))={\rm sgn}(\mu_g-\mu_f)$, and thus $R(r)$ is real everywhere. After solving $\Pi_{(+)}$ and $\Pi_{(g)}$, the variable $\Pi_{(f)}$ is algebraically determined recalling the definition of $\Pi_{(+)}$,
\begin{equation}
 \Pi_{(f)}(t,r)=\frac{K}{r}-\frac{|c|}{\alpha^2\omega}\frac{F_1(t+R(r))}{r}~.
\end{equation}

Finally, in order to explicitly recover the components of the metric perturbations via the relations \eqref{Eq:solution-metric-perturbations}, the solutions for $h_{(-)0}$ and $h_{(-)1}$ can be determined by solving Eqs.~\eqref{Eq:h0minus} and \eqref{Eq:h1minus},
\begin{align}
&\begin{aligned}
     h_{(-)0}(t,r)  =& \; r^2 F_2(t+R(r))+\frac{1}{3}\big(\Pi_{(f)}(t,r)-\omega^2\Pi_{(g)}(t,r)\big)-\frac{r^2}{ \coupling}\Ddot{F}_1(t+R(r))\int\frac{\de r}{r^4 W(r)}~,\label{Eq:solh0minus}
\end{aligned}\\[\jot]
&\begin{aligned}
 h_{(-)1}(t,r)=\;&\frac{Q(r)}{W(r)}h_{(-)0}(t,r)-\frac{\dot{F}_1(t+R(r))}{\coupling \,r^2W(r)}~.\label{Eq:solh1minus}
 \end{aligned}
\end{align}
Similarly to the integration function $C(t)$ in \eqref{Eq:solutionh0g}, the free functions $F_1(t+R(r))$ and $F_2(t+R(r))$ should be fixed consistently with boundary conditions.

\subsection{Physical properties of the solution}\label{Subsec: properties}

\subsubsection{Axial dipolar gravitational waves}

In general, the solutions obtained in the previous section describe propagating dipole radiation. As can be seen, the propagation takes place along the characteristic curves with equation $t+R(r)=constant$, so let us analyze these curves in detail.

In the case with $|c|=\omega$, the function $R(r)$ is proportional to the usual tortoise coordinate $r_*:=r+2\mu_g\ln{|\frac{r}{2\mu_g}-1|}$~. Therefore, the characteristic curves $t+R(r)=constant$ followed by dipolar waves are null, and thus they propagate at the speed of light. An interesting property is that, depending on the sign $s={\rm sgn}(c(\mu_g-\mu_f))$, waves can propagate only in one direction: the waves are purely ingoing for $s=1$, while they are purely outgoing for $s=-1$. This result is consistent with that obtained in Ref.~\cite{Babichev:2015zub}, although in that analysis the signs in \eqref{Eq:solTquadrature} are fixed in such a way that only purely ingoing waves are obtained.

However, for cases with $|c|\neq\omega$, the analysis is not so straightforward. Let us first check the causal character of the characteristic curves. The tangent vector to the curves $t+R(r)=constant$ reads
\begin{equation}
 \xi:=-\frac{Q}{W}\partial_t+\partial_r~,
\end{equation}
and its norm, under the physical metric $g$, is given by
\begin{equation}
 g(\xi,\xi)=(1-\Sigma_g^2 Q^2/W^2)/\Sigma_g
=\frac{|c^2-\omega^2|r}{2|\omega^2\mu_g-c^2\mu_f|+|c^2-\omega^2|r}~,
\end{equation}
while $f(\xi,\xi)=\omega^2g(\xi,\xi)$. As can be seen, for $|c|=\omega$ the norm vanishes, as expected, $\xi$ is lightlike and thus the waves propagate at the speed of light, while, for $|c|\neq\omega$, the norm of $\xi$ is positive for all $r>0$ for both metric sectors. Therefore, for $|c|\neq\omega$, gravitational waves follow spacelike curves with respect to both metrics, and the propagation is superluminal.

Concerning the global qualitative behavior of the curves for cases with $|c|\neq \omega$, we first note that the function $R(r)$ has similar properties to the tortoise coordinate. That is, it is smooth (with nonzero derivative) everywhere, except at the horizon $r=2\mu_g$\footnote{In the following, by {\it horizon} we will always refer
to the horizon of the physical metric $g$ located at $r=2\mu_g$.}, where $2 r W(r)|_{r=2\mu_g}=2 c^2(\mu_g-\mu_f)$, and thus the logarithmic term in \eqref{Eq:integral-I} diverges. More precisely, up to additive constants, around the horizon, $R(r)=s\, r_* +\mathcal{O}\left(e^{\frac{r_*}{2\mu_g}}\right)$, and therefore this function diverges as $R(r)\to-s\infty$ as $r\to 2\mu_g$.

On the one hand, outside the black hole, in the domain $r\in(2\mu_g,\infty)$, the function $R(r)$ is either monotonically decreasing or increasing for $s=-1, 1$, respectively, and it goes as $R(r)\approx s \sqrt{r}$ for large $r$. Therefore, solutions with $s=1$ are purely ingoing, while solutions with $s=-1$ are purely outgoing. On the other hand, inside the black hole $r\in(0,2\mu_g)$, since $R(r)$ is monotonic (increasing or decreasing for $s=-1, 1$, respectively) with $R(0)$ being finite, all the characteristic curves $t+R(r)=constant$ hit the singularity $r=0$ at a certain finite value of $t$ time.

Since the Schwarzschild coordinates $(t,r)$ are singular at the horizon, in order to study the behavior of the waves around that surface, it is convenient to introduce null coordinates $u\coloneqq t-r_*$ and $v\coloneqq t+r_*$. In ingoing Eddington-Finkelstein coordinates $(v,r)$, the characteristic curves read $v+R(r)-r_*=constant$. From the properties of the function $R(r)$, one can show that, for $s=1$, $R(r)-r_*$ is monotonically decreasing for all $r>0$, and thus the characteristic curves go from $r\to+\infty$, as $v\to+ \infty$, to reach $r=0$ at a finite value of $v$ time. However, for $s=-1$, $R(r)-r_*$ diverges at $r=2\mu_g$, and thus the characteristic curves do not cross the black-hole horizon. Conversely, using the outgoing Eddington-Finkelstein coordinates $(u,r)$, one can deduce that purely outgoing solutions $(s=-1)$ do cross the white-hole horizon, while purely ingoing waves $(s=1)$ do not.

\begin{figure}[t]
\centering
   \vspace{-2cm}
 \begin{subfigure}{0.45\textwidth}
 \hspace{-0.8cm}\includegraphics[width=0.85\textwidth, angle=45]{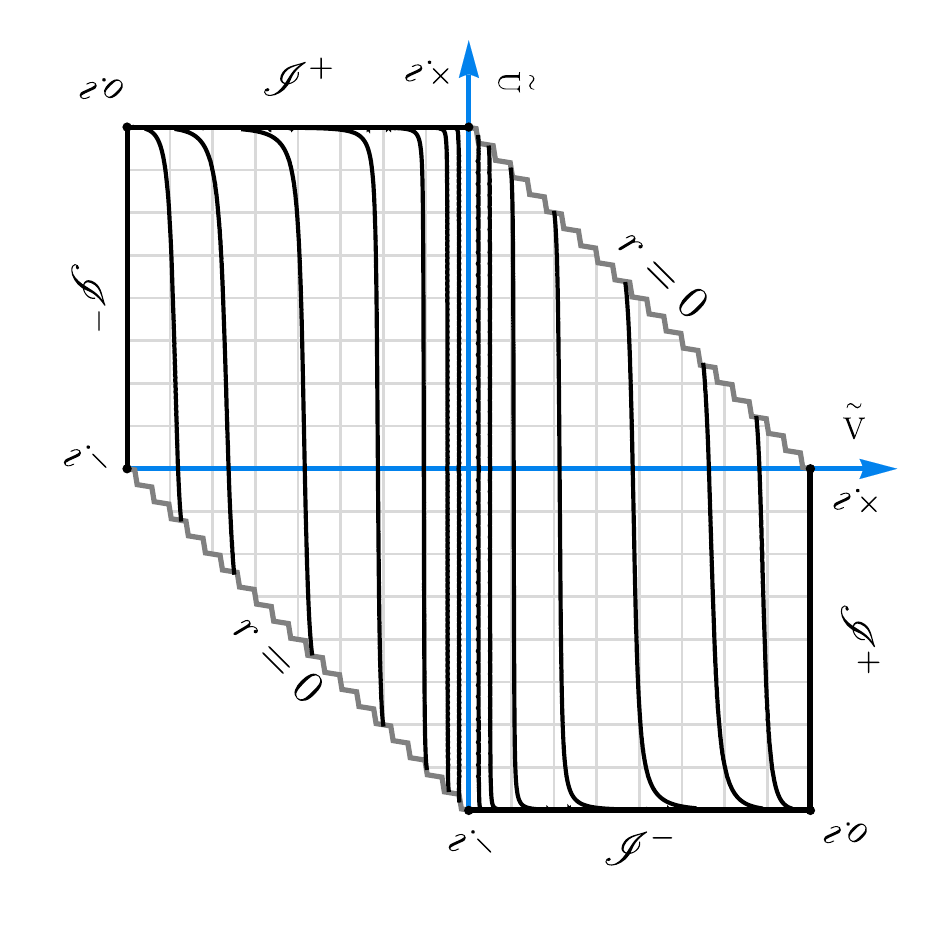}
    \vspace{-3cm}
    \caption{\centering Generic behavior of the characteristic curves for $s=1$, which describe purely ingoing waves.
    The displayed curves correspond to the particular choice of parameters
    $c=1$, $\mu_g=1$, $\mu_f=1/2$, $\omega=7/8$.} \label{fig:diagram-ingoing}
\end{subfigure}
\begin{subfigure}{0.45\textwidth}
        \hspace{-0.8cm} \includegraphics[width=0.85\textwidth, angle=45]{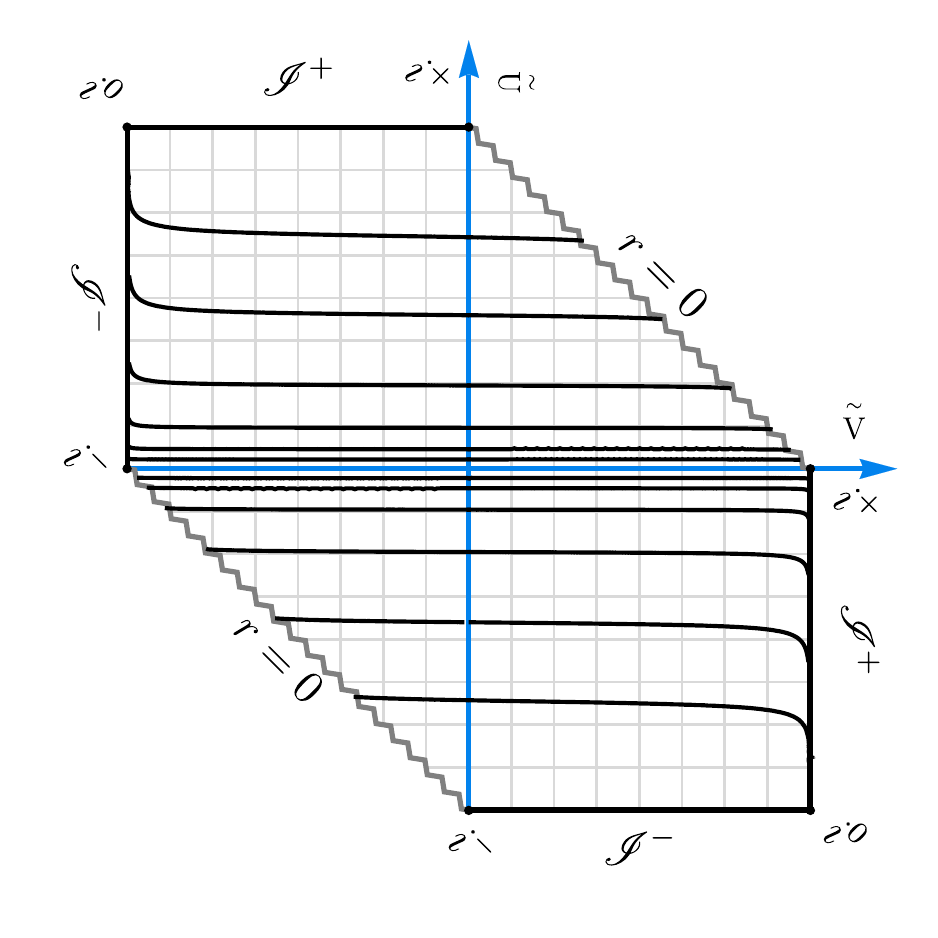}
        \vspace{-3cm}
        \caption{\centering Generic behavior of the characteristic curves for $s=-1$, which describe purely outgoing waves. The displayed curves correspond to the particular choice of parameters $c=1$, $\mu_g=1$, $\mu_f=2$, $\omega=9/8$.}
        \label{fig:diagram-outgoing}
\end{subfigure}
    \caption{The conformal diagrams show the characteristic curves for the propagation of dipole radiation. The direction of propagation is either ingoing or outgoing, depending on the value of the sign factor $s$. In either case, the characteristic curves divide spacetime into two disconnected regions, with each asymptotic region connected either to the black-hole or to the white-hole interior. Even if the characteristic curves are spacelike, no propagation takes place between the two asymptotically flat regions.}
    \label{fig:conformal-diagrams}
\end{figure}

In order to illustrate more clearly the dynamical evolution of these waves, let us represent them in the conformal diagram corresponding to the metric sector $g$. For such a purpose, we consider the Kruskal-Szekeres coordinates, defined as $U\coloneqq -e^{-u/(4 \mu_g)}$ and $V\coloneqq e^{v/(4\mu_g)}$, for which the physical metric $g$ takes the form,
\begin{equation}
 \text{d}s^2_g=-\frac{32 \mu_g^3}{r} e^{-r/(2\mu_g)} \text{d}U \text{d} V+r^2\left(\text{d}\theta^2+\sin^2{\theta}\text{d}\phi^2\right)~,
\end{equation}
and it is thus regular for all $r>0$. By performing the standard compactification $\widetilde{U}\coloneqq {\rm arctan}\left( U \right)$ and $\widetilde{V}\coloneqq {\rm arctan}\left(V \right)$ of these coordinates, we obtain the diagrams shown in Figs.~\ref{fig:diagram-ingoing} and \ref{fig:diagram-outgoing} for the cases with $s=1$ and $s=-1$, respectively. As commented above, for $s=1$ the waves are purely ingoing: they begin at $i^0$, cross the black-hole horizon $r=2\mu_g$, and end up at the singularity $r=0$ in finite time. For $s=-1$ the waves have just the opposite behavior and propagate outward from the white-hole interior to the asymptotically flat region. In all cases, and as already explained above, characteristic curves cross the horizon only once. Thus, even if the waves propagate superluminally and follow spacelike curves, the interior of the black hole is trapped in the sense that, toward the future, no waves can escape, while the interior of the white hole is antitrapped. In particular, this implies that no signal can be exchanged between the two asymptotically flat regions.

In order to understand the shape of the curves in the conformal diagram, one can check the tangent vector $\xi$, which, in the Kruskal-Szekeres coordinates, is given by
\begin{equation}
   \begin{split}
4\mu_g \xi &=\frac{V}{\Sigma_g}\left(1-\frac{Q\Sigma_g}{W}\right) \partial_{V}+\frac{U}{\Sigma_g}\left(1+\frac{Q\Sigma_g}{W}\right) \partial_{U}\\&=\frac{1}{\Sigma_g}\left(1-\frac{Q\Sigma_g}{W}\right) \sin\widetilde V\cos\widetilde V\partial_{\widetilde V}+\frac{1}{\Sigma_g}\left(1+\frac{Q\Sigma_g}{W}\right) \sin\widetilde U\cos\widetilde U\partial_{\widetilde U}~.
\end{split} 
\end{equation}
On the one hand, for large $r$,
\begin{equation}
 \frac{1}{\Sigma_g}\left(1\pm\frac{Q\Sigma_g}{W}\right)=1\pm s |c| \sqrt{\frac{2(\mu_g-\mu_f)}{(c^2-\omega^2)r}}+{\cal O}(1/r)~,
\end{equation}
and thus, in this regime, the tangent vector takes the approximate form,
\begin{equation}
 4\mu_g \xi\approx  V \partial_{V} + U \partial_{U}=
\sin\widetilde V \cos\widetilde V\partial_{\widetilde V}+\sin\widetilde U \cos\widetilde U\partial_{\widetilde U}~.
\end{equation}
Therefore, in the vicinity of the different null infinities, $\xi$ tends to be coparallel either to $\partial_{\widetilde U}$ or $\partial_{\widetilde V}$. For instance, at the $\mathscr{I}^{-}$ defined by $U\to-\infty$, and thus $\widetilde U=-\pi/2$, with $V$ finite, the tangent vector tends to be coparallel to $\partial_{\widetilde V}$, while, around the $\mathscr{I}^{+}$ at $V\to+\infty$, with $U$ finite, the tangent vector tends to be coparallel to $\partial_{\widetilde U}$.

On the other hand, concerning the behavior of the curves around the horizon, taking into account that $\Sigma_g=-2\mu_g  U V e^{-r/(2\mu_g)}/r$, an expansion around $r=2\mu_g$ leads to
\begin{align}\label{nearhorizonexpansion}
\frac{1}{\Sigma_g}\left(1\pm\frac{Q\Sigma_g}{W}\right)&=\frac{1}{\Sigma_g}\left[
1\pm s\mp \frac{s|c^2-\omega^2|}{4 c^2|\mu_g-\mu_f|}(r-2\mu_g)+{\cal O}((r-2\mu_g)^2)\right]
\nonumber\\
&=-\frac{e}{ U V}\left[
1\pm s-\frac{2\mu_g  U  V}{e}\left(\frac{1\pm s}{\mu_g}\mp \frac{s|c^2-\omega^2|}{4 c^2|\mu_g-\mu_f|}\right)+{\cal O}( U^2  V^2)\right].
\end{align}
For definiteness, let us first consider the solutions with $s=1$. From this expression, one then obtains the following approximate form of the vector for the near-horizon regime,
\begin{equation}\label{nearhorizonks1}
 4\mu_g \xi\approx \frac{\mu_g|c^2-\omega^2|}{2 c^2|\mu_g-\mu_f|}  V
 \partial_{ V}-\left[\frac{2e}{ V}-2\mu_g  U \left(\frac{2}{\mu_g}- \frac{|c^2-\omega^2|}{4 c^2|\mu_g-\mu_f|}\right)\right] \partial_{U}~.
 \end{equation}
In fact, this expression is exact at $r=2\mu_g$, and thus, from \eqref{nearhorizonks1}, it is straightforward to see that, when the solutions with $s=1$ cross the black-hole horizon $(U=0)$, the tangent vector takes the form,
\begin{equation}
\xi\big{|}_{U=0}=\frac{|c^2-\omega^2|}{8c^2|\mu_g-\mu_f|}V \partial_{V}-\frac{e}{2\mu_g V}\partial_{U}~.
\end{equation}
Furthermore, these purely ingoing solutions ($s=1$) do not cross the white-hole horizon $(V= 0)$, but, as they approach it, their tangent vector tends to
\begin{equation}\label{nearwhitehorizonks1}
 \xi\to-\frac{e}{2\mu_g V}\partial_{U}~.
\end{equation}
Hence, near the white-hole horizon, characteristic curves tend to the null $V= constant$ curves. Similarly, from \eqref{nearhorizonexpansion} it is straightforward to obtain the near-horizon tangent vector for the solutions with $s=-1$, and conclude that, in the vicinity of the black-hole horizon, which they never cross, the corresponding characteristic curves tend to parallels to the null $U= constant$ curves.
f
We note that, although the characteristic curves are spacelike for $|c|\neq\omega$, this does not lead to closed timelike curves since each characteristic crosses once, and only once, any given $t=constant$ surface at finite values of $r$. Hence, the Cauchy initial value problem for perturbations is well-posed, which is consistent with the considerations made in Ref.~\cite{Hassan:2017ugh} for configurations of the light cones in the background at hand (see Tables~\ref{table:classification-case1-2}, \ref{table:classification-case3} in the \hyperlink{appendix}{Appendix}), and also with the general analysis of Ref.~\cite{Geroch:2010da} on the causal cones of fields propagating superluminal. However, let us consider a static observer at a fixed value of $r>2\mu_g$, outside the horizon, with four-velocity $\Sigma_g^{-1/2} \partial_t$ (normalized with respect to the physical metric $g$). The propagation speed ${\cal V}$ of a pulse following the characteristics, as measured by such a static observer,
is given by (see, e.g., Eq.~8 of Ref. \cite{Crawford:2001jh})
\begin{equation}\label{Eq:velocity}
{\cal V}^2=1+\frac{g(\xi,\xi)}{\left(g(\xi,\Sigma_g^{-1/2} \partial_t)\right)^2}=\frac{W^2}{Q^2\Sigma_g^2}=\frac{|\omega^2-c^2|}{2 c^2|\mu_f-\mu_g|} r+\frac{|c^2\mu_f-\omega^2\mu_g|}{c^2|\mu_f-\mu_g|}.
\end{equation}
As can be seen, $|{\cal V}|$ is a monotonically increasing function of $r$, with $|{\cal V}|\to1$ close to the horizon $r\to2\mu_g$, and $|{\cal V}|\to\infty$ as $r\to\infty$. This is clearly unphysical since there is no reason to expect that gravitational waves are accelerated as they depart from the black hole. Therefore, even if the Cauchy problem is well posed, we conclude that all solutions with $|c|\neq \omega$ are unphysical and should be disregarded. The only background solution that provides a reasonable dynamics of the gravitational waves corresponds to the case with
$|c|=\omega$.

\subsubsection{Static limit}

To complete the analysis of the solution, let us comment on its static limit. In the static case, the solutions to the perturbative dynamics take a particularly simple form
\begin{equation}
    h_{(g)0}=\frac{k_1}{r}+k_2r^2~,\qquad h_{(f)0}=\frac{k_3+\omega^2c_1}{r}+(k_4+\omega^2k_2)r^2~,\qquad h_{(-)1}=\left(\frac{k_3}{r}+k_4r^2\right)\frac{Q}{W}~.
\end{equation}
where $k_1,\,k_2,\,k_3$, and $k_4$ are free integration constants that are related to the above free integration functions as $k_1=-F_1/3,\,k_2=C,\,k_3=K/3-F_1(|c|/(\alpha^2\omega)-\omega^2)/3$, and $k_4=F_2$. In order to ensure that the perturbative approach does not break down at large $r$, we must set $k_2=k_4=0$. Similarly to the general solution obtained earlier in the dynamical case \eqref{Eq:solution-metric-perturbations}, we note that we can only solve the combination $h_{(-)1}$, but not the metric perturbations $h_{(g)1}$ and $h_{(f)1}$ separately.

Now, if we assume that there is just one perturbative dipolar mode, one can always choose the coordinate axes so that its corresponding magnetic number is $m=0$. Then axial symmetry is preserved, and one can compute the usual Komar conserved charges \cite{Jaramillo:2010ay} in each metric sector, i.e., the Komar masses $\mathcal{M}_{i}$ and the Komar angular momenta $J_i$ associated to each perturbed metric $g$ and $f$, which are coordinate independent. The conserved masses are associated to the fact that the spacetimes are stationary and, as expected, are given by
 \begin{equation}
        \mathcal{M}_g=\mu_g\,,\qquad \mathcal{M}_f=\omega\mu_f~.
    \end{equation}
On the other hand, as a result of the axisymmetry of the spacetimes, one has the conserved angular momenta, given by
    \begin{equation}\label{Eq:angular-momenta}
        J_g=-\frac{k_1}{4}\sqrt{\frac{3}{\pi}}\,,\qquad J_f=-\frac{(k_3+\omega^2k_1)\,\omega}{4|c|}\sqrt{\frac{3}{\pi}}~.
    \end{equation}
Note that, although the free constant $c$ appears explicitly in the expression for the angular momentum of the perturbed metric $f$, it is not straightforward to deduce its direct physical implications, since it could be reabsorbed by a redefinition of the integration constants $k_1$ and $k_3$. All in all, we conclude that in the static case perturbed solutions describe slowly rotating black holes in both metric sectors.

\section{Conclusions}\label{Sec: conclusions}

In the framework of the bimetric theory of gravity, we have considered perturbations of the most general spherically symmetric, static, and asymptotically flat vacuum background, such that the two metrics are nonbidiagonal (i.e., there is no coordinate chart in which both metrics are simultaneously diagonal) and they possess the same isometries. This background solution is given by \eqref{Eq:background-line-elements}, along with \eqref{Eq:T_finalform}, and it is completely characterized by four constant parameters. In addition to the two mass parameters $\mu_g$ and $\mu_f$, and the ratio between the areal radii of the two metrics $\omega$, there is another free parameter $c$, which represents the proportionality factor between the (appropriately normalized) time-translation invariance Killing vector field of each metric. However, these four constants are not completely arbitrary. Rather, as studied in detail in Sec.~\ref{Sec: background}, they must satisfy certain specific relations to ensure that the solution is smooth and well-defined.

Even if previous perturbative studies have been carried out on such background, the specific value $|c|=\omega$ has always been imposed at the outset. Therefore, the main focus of the present work has been to understand the physical implications of allowing generic values for the parameter $c$. For this purpose, we have considered the dynamics of a dipolar $(l=1)$ perturbative mode with axial (sometimes also known as odd) polarity. Although this is a simple setup, contrary to GR, the linearized bimetric equations of motion imply nontrivial dynamics for this mode. We have been able to obtain the full analytic solution for such equations for any gauge, which has allowed us to determine in detail the behavior of the dipolar radiation. In particular, we have found that the radiation is either purely ingoing or purely outgoing, depending on the sign $s={\rm sgn}(c(\mu_g-\mu_f))$. Furthermore, and more importantly, for $|c|\neq\omega$, we have shown that the dipolar radiation propagates along spacelike characteristic curves with respect to both metrics, and thus the propagation is superluminal. It is only for $|c|=\omega$ that the characteristic curves are lightlike. In particular, for $|c|=\omega$, the results obtained are consistent with those presented in Ref.~\cite{Babichev:2015zub}, with the difference that, since their analysis corresponds to a specific choice of the sign $s$, their solution for dipolar axial perturbations describes purely ingoing waves.

It is expected that bimetric gravity admits superluminality, possibly without violating causality \cite{Hassan:2017ugh}, with the causal cones given by the convex hull of the null cones of the two metrics \cite{Kocic:2018yvr}. In fact, in the present case, the solutions are well-behaved in the sense that the Cauchy problem is well-posed since no closed causal curves arise, as illustrated in Fig.~\ref{fig:conformal-diagrams}. Furthermore, purely ingoing (outgoing) solutions cross once, and only once, the black-hole (white-hole) horizon. Therefore, despite the superluminal propagation, waves cannot escape from the interior of the black hole, and no propagation takes place between the two asymptotically flat ends.

Nevertheless, as can be seen from \eqref{Eq:velocity}, unless $|c|=\omega$, the propagation speed of a pulse that follows the characteristic curves obtained for the dipolar waves, as measured by a static observer, grows monotonically with $r$. It is only for $|c|=\omega$ that the velocity equals the speed of light. Based on this result, we conclude that only background solutions with $|c|=\omega$ are physically meaningful, thus providing strong theoretical support for this choice.

The propagation of the axial dipolar mode is a distinctive feature of bimetric gravity, while in general relativity this mode is nonradiative and corresponds to the angular momentum of a slowly rotating black hole. Thus, quantifying the excitation of this mode in astrophysical processes (e.g., binary mergers) may help place observational bounds on the parameters of the theory. This investigation is left for future work.
We also note that Ref.~\cite{Cardoso:2023dwz} quantified the excitation of the polar $l=1$ mode by extreme mass ratio inspirals (EMRIs) in massive gravity. Although these bounds do not directly apply to bimetric gravity (as massive gravity is only recovered in the $M_f/M_g\to\infty$ limit), it would be interesting to perform a similar analysis in more general regimes of bimetric gravity---particularly close to the limit $M_f/M_g\ll1$, which corresponds to general relativity.

\section*{Acknowledgements}
We thank Marcus H{\"o}g{\aa}s for helpful comments on a previous version of the manuscript.
ASO acknowledges Nordita for hospitality, where part of this work was carried out.
ASO also acknowledges financial support from the fellowship PIF21/237
of the UPV/EHU. MdC acknowledges support from INFN iniziativa specifica GeoSymQFT. 
This work has been supported by the Basque Government Grant
\mbox{IT1628-22} and by the Grant PID2021-123226NB-I00 (funded by
MCIN/AEI/10.13039/501100011033 and by ``ERDF A way of making Europe'').

\appendix

\section*{Appendix: \hypertarget{appendix}{Configurations of the light cones}}\label{App:cones}

In this appendix, we examine the different possible types of bimetric configurations for the solution \eqref{Eq:background-line-elements} along with \eqref{Eq:T_finalform}, classified in terms of the intersections of the two null cones. Tables \ref{table:classification-case1-2} and \ref{table:classification-case3} show the possible Types of local bimetric configurations corresponding to the line elements \eqref{Eq:metric-g}, \eqref{Eq:metric-f}, with $T_I$ given by \eqref{Eq:T_finalform}, based on the general classification of Ref.~\cite{ Hassan:2017ugh} and computed using the results in Ref.~\cite{Kocic:2019ahm}. The quantity $r_0$ is defined as $r_0\coloneqq\frac{2}{\omega^2}\left(\omega^2\mu_g-c^2(\mu_f-\mu_g)\right)$. Clearly, when $r_0<0$ some of the configurations listed in the table are not accessible. The descriptors in brackets (``$f$ inside of $g$'' or ``$g$ inside of $f$'') refer to the relative orientation of the light cones of the two metrics, and should not be confused with the relative position of their respective horizons (which is uniquely determined in each case). On the other hand, the label (L) (respectively (R)) in Table \ref{table:classification-case3} indicates that both metrics have the left (respectively right) null direction in common. We observe that all configurations listed below ensure that the two metrics admit compatible 3+1 decompositions, which is a necessary condition for the initial value problem to be well-posed in the bimetric theory, cf.~Ref.~\cite{Hassan:2017ugh}.

\begin{table}[H]
    \centering
\begin{minipage}{0.48\textwidth}
\centering
\begin{tabular}{|c|c|c|}
 \hline
\multicolumn{3}{|c|}{$\boldsymbol{\omega<|c|}$ and $\boldsymbol{c^2\mu_f\leq \omega^2\mu_g}$}                  \\ \hline \hline
 & $c>0$            & $c<0$           \\ \hline
$r<2\mu_g$      & I ($f$ inside $g$)    & I ($f$ inside $g$)    \\ \hline
$2\mu_g<r<r_0$  & IIb ($f$ left of $g$) & IIb ($g$ left of $f$) \\ \hline
$r>r_0$         & I ($g$ inside $f$)    & I ($g$ inside $f$)    \\ \hline
\end{tabular}
        \end{minipage}%
    \hfill
    \begin{minipage}{0.48\textwidth}
        \centering
        \begin{tabular}{|c|c|c|}
        \hline
\multicolumn{3}{|c|}{$\boldsymbol{\omega>|c|}$ and $\boldsymbol{c^2\mu_f\geq \omega^2\mu_g}$}                  \\ \hline \hline
 & $c>0$            & $c<0$           \\ \hline
$0<r<r_0$    & I ($g$ inside $f$)    & I ($g$ inside $f$)    \\ \hline
$r_0<r<2\mu_g$  & IIb ($g$ left of $f$) & IIb ($f$ left of $g$) \\ \hline
$r>2\mu_g$      & I ($f$ inside $g$)    & I ($f$ inside $g$)    \\ \hline
\end{tabular}
    \end{minipage}
    \caption{Classification of the causal types for different consistent choices of the background parameters.}
    \label{table:classification-case1-2}
\end{table}

\begin{table}[H]
\centering
\begin{tabular}{|cccccc|}
\hline
\multicolumn{6}{|c|}{$\boldsymbol{|c|=\omega}$}                  \\ \hline \hline
\multicolumn{1}{|c|}{$\boldsymbol{\mu_g>\mu_f}$} & \multicolumn{1}{c|}{$c>0$}          & \multicolumn{1}{c|}{$c<0$}         & \multicolumn{1}{||c|}{$\boldsymbol{\mu_g<\mu_f}$} & \multicolumn{1}{c|}{$c>0$}          & $c<0$         \\ \hline
\multicolumn{1}{|c|}{$r<2\mu_g$}             & \multicolumn{1}{c|}{IIa (L)}               & \multicolumn{1}{c|}{IIa (R)}               & \multicolumn{1}{||c|}{$0<r<r_0$}              & \multicolumn{1}{c|}{IIa (R)}               & IIa (L)               \\ \hline
\multicolumn{1}{|c|}{$2\mu_g<r<r_0$}         & \multicolumn{1}{c|}{IIb ($f$ left of $g$)} & \multicolumn{1}{c|}{IIb ($g$ left of $f$)} & \multicolumn{1}{||c|}{$r_0<r<2\mu_g$}         & \multicolumn{1}{c|}{IIb ($g$ left of $f$)} & IIb ($f$ left of $g$) \\ \hline
\multicolumn{1}{|c|}{$r>r_0$}                & \multicolumn{1}{c|}{IIa (L)}               & \multicolumn{1}{c|}{IIa (R)}               & \multicolumn{1}{||c|}{$r>2\mu_g$}             & \multicolumn{1}{c|}{IIa (R)}               & IIa (L)               \\ \hline
\end{tabular}
\caption{Classification of the causal types for backgrounds with $|c|=\omega$.}
\label{table:classification-case3}
\end{table}

\bibliographystyle{bib-style}
\bibliography{references}

\providecommand{\href}[2]{#2}\begingroup\raggedright\begin{thebibliography}{10}

\bibitem{Hassan:2011vm}
S.~F. Hassan and R.~A. Rosen, ``{On Non-Linear Actions for Massive Gravity},'' JHEP {\bf 07} (2011) 009, \href{http://arXiv.org/abs/1103.6055}{{\tt arXiv:1103.6055}}.

\bibitem{Hassan:2011tf}
S.~F. Hassan, R.~A. Rosen, and A.~Schmidt-May, ``{Ghost-free Massive Gravity with a General Reference Metric},'' JHEP {\bf 02} (2012) 026, \href{http://arXiv.org/abs/1109.3230}{{\tt arXiv:1109.3230}}.

\bibitem{Volkov:2011an}
M.~S. Volkov, ``{Cosmological solutions with massive gravitons in the bigravity theory},'' JHEP {\bf 01} (2012) 035, \href{http://arXiv.org/abs/1110.6153}{{\tt arXiv:1110.6153}}.

\bibitem{Volkov:2012zb}
M.~S. Volkov, ``{Exact self-accelerating cosmologies in the ghost-free massive gravity -- the detailed derivation},'' Phys. Rev. D {\bf 86} (2012) 104022, \href{http://arXiv.org/abs/1207.3723}{{\tt arXiv:1207.3723}}.

\bibitem{DeFelice:2014nja}
A.~De~Felice, A.~E. G\"umr\"uk\c{c}\"uo\u{g}lu, S.~Mukohyama, N.~Tanahashi, and T.~Tanaka, ``{Viable cosmology in bimetric theory},'' JCAP {\bf 06} (2014) 037, \href{http://arXiv.org/abs/1404.0008}{{\tt arXiv:1404.0008}}.

\bibitem{Akrami:2015qga}
Y.~Akrami, S.~F. Hassan, F.~K\"onnig, A.~Schmidt-May, and A.~R. Solomon, ``{Bimetric gravity is cosmologically viable},'' Phys. Lett. B {\bf 748} (2015) 37--44, \href{http://arXiv.org/abs/1503.07521}{{\tt arXiv:1503.07521}}.

\bibitem{Aoki:2014cla}
K.~Aoki and K.-i. Maeda, ``{Dark matter in ghost-free bigravity theory: From a galaxy scale to the universe},'' Phys. Rev. D {\bf 90} (2014) 124089, \href{http://arXiv.org/abs/1409.0202}{{\tt arXiv:1409.0202}}.

\bibitem{Bernard:2014psa}
L.~Bernard and L.~Blanchet, ``{Phenomenology of Dark Matter via a Bimetric Extension of General Relativity},'' Phys. Rev. D {\bf 91} (2015), no.~10, 103536, \href{http://arXiv.org/abs/1410.7708}{{\tt arXiv:1410.7708}}.

\bibitem{Blanchet:2015sra}
L.~Blanchet and L.~Heisenberg, ``{Dark Matter via Massive (bi-)Gravity},'' Phys. Rev. D {\bf 91} (2015) 103518, \href{http://arXiv.org/abs/1504.00870}{{\tt arXiv:1504.00870}}.

\bibitem{Blanchet:2015bia}
L.~Blanchet and L.~Heisenberg, ``{Dipolar Dark Matter with Massive Bigravity},'' JCAP {\bf 12} (2015) 026, \href{http://arXiv.org/abs/1505.05146}{{\tt arXiv:1505.05146}}.

\bibitem{Babichev:2016hir}
E.~Babichev, L.~Marzola, M.~Raidal, A.~Schmidt-May, F.~Urban, H.~Veerm\"ae, and M.~von Strauss, ``{Bigravitational origin of dark matter},'' Phys. Rev. D {\bf 94} (2016), no.~8, 084055, \href{http://arXiv.org/abs/1604.08564}{{\tt arXiv:1604.08564}}.

\bibitem{Dwivedi:2024okk}
S.~Dwivedi and M.~H\"og\r{a}s, ``{2D BAO vs. 3D BAO: Solving the Hubble Tension with Bimetric Cosmology},'' Universe {\bf 10} (2024), no.~11, 406, \href{http://arXiv.org/abs/2407.04322}{{\tt arXiv:2407.04322}}.

\bibitem{vonStrauss:2011mq}
M.~von Strauss, A.~Schmidt-May, J.~Enander, E.~Mortsell, and S.~F. Hassan, ``{Cosmological Solutions in Bimetric Gravity and their Observational Tests},'' JCAP {\bf 03} (2012) 042, \href{http://arXiv.org/abs/1111.1655}{{\tt arXiv:1111.1655}}.

\bibitem{Caravano:2021aum}
A.~Caravano, M.~L\"uben, and J.~Weller, ``{Combining cosmological and local bounds on bimetric theory},'' JCAP {\bf 09} (2021) 035, \href{http://arXiv.org/abs/2101.08791}{{\tt arXiv:2101.08791}}.

\bibitem{Hogas:2021lns}
M.~H\"og\r{a}s and E.~M\"ortsell, ``{Constraints on bimetric gravity. Part II. Observational constraints},'' JCAP {\bf 05} (2021) 002, \href{http://arXiv.org/abs/2101.08795}{{\tt arXiv:2101.08795}}.

\bibitem{Hogas:2021saw}
M.~H\"og\r{a}s and E.~M\"ortsell, ``{Constraints on bimetric gravity from Big Bang nucleosynthesis},'' JCAP {\bf 11} (2021) 001, \href{http://arXiv.org/abs/2106.09030}{{\tt arXiv:2106.09030}}.

\bibitem{Hogas:2021fmr}
M.~H\"og\r{a}s and E.~M\"ortsell, ``{Constraints on bimetric gravity. Part I. Analytical constraints},'' JCAP {\bf 05} (2021) 001, \href{http://arXiv.org/abs/2101.08794}{{\tt arXiv:2101.08794}}.

\bibitem{Hogas:2022owf}
M.~H\"og\r{a}s, {\em {Was Einstein Wrong? : Theoretical and observational constraints on massive gravity}}.
\newblock PhD thesis, Stockholm University, Faculty of Science, Department of Physics., 2022.

\bibitem{Babichev:2015xha}
E.~Babichev and R.~Brito, ``{Black holes in massive gravity},'' Class. Quant. Grav. {\bf 32} (2015) 154001, \href{http://arXiv.org/abs/1503.07529}{{\tt arXiv:1503.07529}}.

\bibitem{Torsello:2017cmz}
F.~Torsello, M.~Kocic, and E.~Mortsell, ``{Classification and asymptotic structure of black holes in bimetric theory},'' Phys. Rev. D {\bf 96} (2017), no.~6, 064003, \href{http://arXiv.org/abs/1703.07787}{{\tt arXiv:1703.07787}}.

\bibitem{Comelli:2011wq}
D.~Comelli, M.~Crisostomi, F.~Nesti, and L.~Pilo, ``{Spherically Symmetric Solutions in Ghost-Free Massive Gravity},'' Phys. Rev. D {\bf 85} (2012) 024044, \href{http://arXiv.org/abs/1110.4967}{{\tt arXiv:1110.4967}}.

\bibitem{Babichev:2014tfa}
E.~Babichev and A.~Fabbri, ``{Rotating black holes in massive gravity},'' Phys. Rev. D {\bf 90} (2014) 084019, \href{http://arXiv.org/abs/1406.6096}{{\tt arXiv:1406.6096}}.

\bibitem{Babichev:2014fka}
E.~Babichev and A.~Fabbri, ``{A class of charged black hole solutions in massive (bi)gravity},'' JHEP {\bf 07} (2014) 016, \href{http://arXiv.org/abs/1405.0581}{{\tt arXiv:1405.0581}}.

\bibitem{Babichev:2013una}
E.~Babichev and A.~Fabbri, ``{Instability of black holes in massive gravity},'' Class. Quant. Grav. {\bf 30} (2013) 152001, \href{http://arXiv.org/abs/1304.5992}{{\tt arXiv:1304.5992}}.

\bibitem{Brito:2013wya}
R.~Brito, V.~Cardoso, and P.~Pani, ``{Massive spin-2 fields on black hole spacetimes: Instability of the Schwarzschild and Kerr solutions and bounds on the graviton mass},'' Phys. Rev. D {\bf 88} (2013), no.~2, 023514, \href{http://arXiv.org/abs/1304.6725}{{\tt arXiv:1304.6725}}.

\bibitem{Babichev:2014oua}
E.~Babichev and A.~Fabbri, ``{Stability analysis of black holes in massive gravity: a unified treatment},'' Phys. Rev. D {\bf 89} (2014), no.~8, 081502, \href{http://arXiv.org/abs/1401.6871}{{\tt arXiv:1401.6871}}.

\bibitem{Brito:2013xaa}
R.~Brito, V.~Cardoso, and P.~Pani, ``{Black holes with massive graviton hair},'' Phys. Rev. D {\bf 88} (2013) 064006, \href{http://arXiv.org/abs/1309.0818}{{\tt arXiv:1309.0818}}.

\bibitem{Gervalle:2020mfr}
R.~Gervalle and M.~S. Volkov, ``{Asymptotically flat hairy black holes in massive bigravity},'' Phys. Rev. D {\bf 102} (2020), no.~12, 124040, \href{http://arXiv.org/abs/2008.13573}{{\tt arXiv:2008.13573}}.

\bibitem{Babichev:2015zub}
E.~Babichev, R.~Brito, and P.~Pani, ``{Linear stability of nonbidiagonal black holes in massive gravity},'' Phys. Rev. D {\bf 93} (2016), no.~4, 044041, \href{http://arXiv.org/abs/1512.04058}{{\tt arXiv:1512.04058}}.

\bibitem{Wood:2024acv}
K.~Wood, P.~M. Saffin, and A.~Avgoustidis, ``{Black holes in multimetric gravity},'' Phys. Rev. D {\bf 109} (2024), no.~12, 124006, \href{http://arXiv.org/abs/2402.17835}{{\tt arXiv:2402.17835}}.

\bibitem{Wood:2024eol}
K.~Wood, P.~M. Saffin, and A.~Avgoustidis, ``{Black holes in multimetric gravity. II. Hairy solutions and linear stability of the non- and partially proportional branches},'' Phys. Rev. D {\bf 111} (2025), no.~2, 024057, \href{http://arXiv.org/abs/2410.10976}{{\tt arXiv:2410.10976}}.

\bibitem{Volkov:2012wp}
M.~S. Volkov, ``{Hairy black holes in the ghost-free bigravity theory},'' Phys. Rev. D {\bf 85} (2012) 124043, \href{http://arXiv.org/abs/1202.6682}{{\tt arXiv:1202.6682}}.

\bibitem{Volkov2015}
M.~S. Volkov, {\em Hairy Black Holes in Theories with Massive Gravitons}, pp.~161--180.
\newblock Springer International Publishing, Cham, 2015.

\bibitem{Torsello:2017ouh}
F.~Torsello, M.~Kocic, M.~H\"og\r{a}s, and E.~Mortsell, ``{Spacetime symmetries and topology in bimetric relativity},'' Phys. Rev. D {\bf 97} (2018), no.~8, 084022, \href{http://arXiv.org/abs/1710.06434}{{\tt arXiv:1710.06434}}.

\bibitem{Brizuela:2024smr}
D.~Brizuela, M.~de~Cesare, and A.~Soler~Oficial, ``{Perturbations of bimetric gravity on most general spherically symmetric spacetimes},'' Phys. Rev. D {\bf 109} (2024), no.~12, 124060, \href{http://arXiv.org/abs/2402.15327}{{\tt arXiv:2402.15327}}.

\bibitem{Hassan:2011zd}
S.~F. Hassan and R.~A. Rosen, ``{Bimetric Gravity from Ghost-free Massive Gravity},'' JHEP {\bf 02} (2012) 126, \href{http://arXiv.org/abs/1109.3515}{{\tt arXiv:1109.3515}}.

\bibitem{Hassan:2011hr}
S.~F. Hassan and R.~A. Rosen, ``{Resolving the Ghost Problem in non-Linear Massive Gravity},'' Phys. Rev. Lett. {\bf 108} (2012) 041101, \href{http://arXiv.org/abs/1106.3344}{{\tt arXiv:1106.3344}}.

\bibitem{Bernard:2015mkk}
L.~Bernard, C.~Deffayet, and M.~von Strauss, ``{Massive graviton on arbitrary background: derivation, syzygies, applications},'' JCAP {\bf 06} (2015) 038, \href{http://arXiv.org/abs/1504.04382}{{\tt arXiv:1504.04382}}.

\bibitem{deRham:2014naa}
C.~de~Rham, L.~Heisenberg, and R.~H. Ribeiro, ``{On couplings to matter in massive (bi-)gravity},'' Class. Quant. Grav. {\bf 32} (2015) 035022, \href{http://arXiv.org/abs/1408.1678}{{\tt arXiv:1408.1678}}.

\bibitem{Yamashita:2014fga}
Y.~Yamashita, A.~De~Felice, and T.~Tanaka, ``{Appearance of Boulware\textendash{}Deser ghost in bigravity with doubly coupled matter},'' Int. J. Mod. Phys. D {\bf 23} (2014) 1443003, \href{http://arXiv.org/abs/1408.0487}{{\tt arXiv:1408.0487}}.

\bibitem{Schmidt-May:2015vnx}
A.~Schmidt-May and M.~von Strauss, ``{Recent developments in bimetric theory},'' J. Phys. A {\bf 49} (2016), no.~18, 183001, \href{http://arXiv.org/abs/1512.00021}{{\tt arXiv:1512.00021}}.

\bibitem{Gerlach:1979rw}
U.~H. Gerlach and U.~K. Sengupta, ``{Gauge invariant perturbations on most general spherically symmetric space-times},'' Phys. Rev. D {\bf 19} (1979) 2268--2272.

\bibitem{Hassan:2017ugh}
S.~F. Hassan and M.~Kocic, ``{On the local structure of spacetime in ghost-free bimetric theory and massive gravity},'' JHEP {\bf 05} (2018) 099, \href{http://arXiv.org/abs/1706.07806}{{\tt arXiv:1706.07806}}.

\bibitem{Geroch:2010da}
R.~Geroch, ``{Faster Than Light?},'' AMS/IP Stud. Adv. Math. {\bf 49} (2011) 59--70, \href{http://arXiv.org/abs/1005.1614}{{\tt arXiv:1005.1614}}.

\bibitem{Crawford:2001jh}
P.~Crawford and I.~Tereno, ``{Generalized observers and velocity measurements in general relativity},'' Gen. Rel. Grav. {\bf 34} (2002) 2075, \href{http://arXiv.org/abs/gr-qc/0111073}{{\tt arXiv:gr-qc/0111073}}.

\bibitem{Jaramillo:2010ay}
J.~L. Jaramillo and E.~Gourgoulhon, ``{Mass and Angular Momentum in General Relativity},'' Fundam. Theor. Phys. {\bf 162} (2011) 87--124, \href{http://arXiv.org/abs/1001.5429}{{\tt arXiv:1001.5429}}.

\bibitem{Kocic:2018yvr}
M.~Kocic, ``{Causal propagation of constraints in bimetric relativity in standard 3+1 form},'' JHEP {\bf 10} (2019) 219, \href{http://arXiv.org/abs/1804.03659}{{\tt arXiv:1804.03659}}.

\bibitem{Cardoso:2023dwz}
V.~Cardoso, F.~Duque, A.~Maselli, and D.~Pere\~niguez, ``{Constraints on massive gravity from dipolar mode excitations},'' Phys. Rev. D {\bf 108} (2023), no.~12, 124003, \href{http://arXiv.org/abs/2304.01252}{{\tt arXiv:2304.01252}}.

\bibitem{Kocic:2019ahm}
M.~Kocic, ``{Note on bimetric causal diagrams},'' \href{http://arXiv.org/abs/1904.10407}{{\tt arXiv:1904.10407}}.

\end{thebibliography}\endgroup

\end{document}